\documentclass[%
 aip,
 jcp,
 amsmath,amssymb,
 reprint,%
]{revtex4-1}

\usepackage{graphicx}% Include figure files
\usepackage{dcolumn}% Align table columns on decimal point
\usepackage{bm}% bold math
\usepackage[utf8]{inputenc}
\usepackage[T1]{fontenc}
\usepackage{mathptmx}

\begin{document}

\preprint{AIP/123-QED}

\title{On the calculation of the bandgap of periodic solids with MGGA functionals
using the total energy}

\author{Fabien Tran}
\affiliation{Institute of Materials Chemistry, Vienna University of Technology,
Getreidemarkt 9/165-TC, A-1060 Vienna, Austria}
\author{Jan Doumont}
\affiliation{Institute of Materials Chemistry, Vienna University of Technology,
Getreidemarkt 9/165-TC, A-1060 Vienna, Austria}
\author{Peter Blaha}
\affiliation{Institute of Materials Chemistry, Vienna University of Technology,
Getreidemarkt 9/165-TC, A-1060 Vienna, Austria}
\author{Miguel A. L. Marques}
\affiliation{Institut f\"{u}r Physik, Martin-Luther-Universit\"{a}t Halle-Wittenberg,
D-06099 Halle, Germany}
\author{Silvana Botti}
\affiliation{Institut f\"ur Festk\"orpertheorie und -optik,
Friedrich-Schiller-Universit\"{a}t Jena and European Theoretical Spectroscopy Facility,
Max-Wien-Platz 1, 07743 Jena, Germany}
\author{Albert P. Bart\'{o}k}
\affiliation{Rutherford Appleton Laboratory, Scientific Computing Department Science and
Technology Facilities Council, Didcot OX11 0QX, United Kingdom}
\affiliation{Department of Physics and Warwick Centre for Predictive Modelling,
School of Engineering, University of Warwick, Coventry, CV4 7AL, United Kingdom}

\date{\today}

\begin{abstract}

During the last few years, it has become more and more clear that functionals
of the meta generalized gradient approximation (MGGA) are more accurate
than GGA functionals for the geometry and energetics of electronic systems.
However, MGGA functionals are also potentially more interesting for the
electronic structure, in particular when the potential is non-multiplicative
(i.e., when MGGAs are implemented in the generalized Kohn-Sham framework),
which may help to get more accurate bandgaps.
Here, we show that the calculation of bandgap of solids with MGGA functionals
can be done very accurately also in a non-self-consistent manner.
This scheme uses only the total energy and can, therefore, be very useful when the
self-consistent implementation of a particular MGGA functional is not available.
Since self-consistent MGGA calculations may be difficult to converge,
the non-self-consistent scheme may also help to speed-up the calculations.
Furthermore, it can be applied to any other types of functionals, for which
the implementation of the corresponding potential is not trivial.

\end{abstract}

\maketitle

In density functional theory (DFT)\cite{HohenbergPR64} implemented
using an auxiliary system of noninteracting electrons, either within
the Kohn-Sham (KS)\cite{KohnPR65} or generalized KS (gKS)\cite{SeidlPRB96}
framework, the difference between
the energies of the highest occupied (HO) and lowest unoccupied (LU) orbitals,
\begin{equation}
E_{\text{g}}^{\text{(g)KS}}=\varepsilon_{\text{LU}}-\varepsilon_{\text{HO}},
\label{eq:Eg1}
\end{equation}
is often used to calculate the fundamental (photoemission) bandgap
$E_{\text{g}}$. It is defined formally as ($N$ is the number of electrons
in the system)
\begin{eqnarray}
E_{\text{g}} & = & I(N) - A(N) \nonumber \\
& = & [E_{\text{tot}}(N-1)-E_{\text{tot}}(N)] -
[E_{\text{tot}}(N)-E_{\text{tot}}(N+1)],
\label{eq:Eg2}
\end{eqnarray}
where $I$ and $A$ are the ionization potential and electron affinity, respectively.
However, within the KS framework,\cite{KohnPR65} i.e., with a
multiplicative exchange-correlation potential
$v_{\text{xc}}=\delta E_{\text{xc}}/\delta\rho$,
$E_{\text{g}}^{\text{KS}}$ and
$E_{\text{g}}$ differ:
\begin{eqnarray}
E_{\text{g}} & = & I(N) - A(N) = - \epsilon_{\text{HO}}(N) - [-\epsilon_{\text{HO}}(N+1)] \nonumber \\
& = & \underbrace{\epsilon_{\text{LU}}(N) - \epsilon_{\text{HO}}(N)}_{E_{\text{g}}^{\text{KS}}} +
\underbrace{\epsilon_{\text{HO}}(N+1) - \epsilon_{\text{LU}} (N)}_{\Delta_{\text{xc}}} \nonumber\\
& = & E_{\text{g}}^{\text{KS}} + \Delta_{\text{xc}},
\label{eq:deltaxc} 
\end{eqnarray}
where $\Delta_{\text{xc}}$ is the derivative discontinuity.\cite{PerdewPRL82,ShamPRL83}
As a consequence, with most functionals, either the (unknown) exact one or a
standard approximation of the local density approximation (LDA) or
generalized gradient approximation (GGA),\cite{PerdewPRL96,BeckePRA88,LeePRB88}
$\varepsilon_{\text{LU}}-\varepsilon_{\text{HO}}$
is (much) smaller than the value of $E_{\text{g}}$
obtained from experiment\cite{HeydJCP05} (note that very often
the experimental bandgap is obtained from optical experiment, which however
is direct and includes the excitonic effect).
Nevertheless, a few points should be mentioned. First,
for a periodic solid it was proven that $\Delta_{\text{xc}}=0$ with LDA/GGA
\cite{KraislerJCP14,GorlingPRB15}
(but $\varepsilon_{\text{LU}}-\varepsilon_{\text{HO}}$ is still smaller than
$E_{\text{g}}$). Second, the self-interaction error (SIE)\cite{PerdewPRB81}
(or delocalization error\cite{MoriSanchezPRL08})
may also worsen the discrepancy between $E_{\text{g}}$ and
$E_{\text{g}}^{\text{KS}}$, and actually the absence of a derivative discontinuity and
the SIE are difficult to disentangle.\cite{PerdewAQC90}
Third, within KS-DFT there exist specialized functionals
that are able to provide values of
$\varepsilon_{\text{LU}}-\varepsilon_{\text{HO}}$ close to
$E_{\text{g}}$.\cite{TranPRL09,ArmientoPRL13,VermaJPCL17}

Within the gKS theory,\cite{SeidlPRB96} which concerns
functionals $ E_{\text{xc}}$ that are not explicit functionals of the electron
density $\rho$, $\Delta_{\text{xc}}$ (or a part of it) is included in
$E_{\text{g}}^{\text{gKS}}=\varepsilon_{\text{LU}}-\varepsilon_{\text{HO}}$
[i.e., Eq.~(\ref{eq:deltaxc}) does not hold in gKS theory].
Meta-GGA (MGGA)\cite{DellaSalaIJQC16} and hybrid functionals\cite{BeckeJCP93b}
are usually implemented with a gKS Hamiltonian and should in principle be able
to provide bandgaps that are more accurate than with GGA. This also means
that a direct comparison of $E_{\text{g}}^{\text{gKS}}$ and $E_{\text{g}}$ is
more justified than in the KS theory.\cite{KuemmelRMP08,YangJCP12,YangPRB16,PerdewPNAS17}

MGGA methods are very attractive since they are of the semilocal type,
therefore computationally efficient, and they have shown to be overall
more accurate than GGA functionals for the geometry and energetics of
molecules and solids. This is for instance the case with the recent SCAN
functional,\cite{SunPRL15} which has attracted a lot of attention, see e.g.,
Refs.~\onlinecite{IsaacsPRM18,ZhangNPJCM18} (in passing we note that SCAN
reduces the SIE,\cite{LanePRB18,VarignonPRB19} but on the other hand leads
to over-localization and therefore too large magnetic moments in itinerant
metals\cite{IsaacsPRM18,JanaJCP18a,FuPRL18,MejiaRodriguezPRB19}).
Thus, MGGA functionals are very interesting and promising.

The focus of the present work is on the calculation of the bandgap of solids
with MGGA functionals. We will show that the bandgap can be calculated
very accurately non-self-consistently using the total energy with
Eq.~(\ref{eq:Eg2}). This procedure is very useful when a self-consistent
implementation of MGGAs\cite{NeumannMP96} is not available or the
computational effort needs to be reduced (MGGAs can be notably more expensive than
GGAs\cite{BienvenuJCTC18,MejiaRodriguezPRB18} and may require
more iterations to achieve self-consistent field convergence).
Thus, the proposed scheme may also be very helpful to apply MGGAs
more efficiently in applications involving very large systems or
for high-throughput materials screening.

Contrary to the common belief\cite{MoralesGarciaJPCC17,ChenPRM18} that using
Eq.~(\ref{eq:Eg2}) for the calculation of the bandgap of periodic systems
is technically difficult and poses problems, it has been recently underlined in
Refs.~\onlinecite{GorlingPRB15,TrushinPRB16,PerdewPNAS17}
(see also a related discussion in Ref.~\onlinecite{VlcekJCP15}) that it is not
the case. Here, Eq.~(\ref{eq:Eg2}) is used to calculate $E_{\text{g}}$ with
MGGA functionals. Briefly, $E_{\text{tot}}(N-1)$, the total energy of the
whole solid which consists of $N_{\bm{k}}$ unit cells (i.e., the number of
$\bm{k}$ points in the first Brillouin zone used in the calculation) and $N$
electrons is evaluated with the electron density
$\rho^{N-1}=\rho^{N}-\left(1/N_{\bm{k}}\right)\left\vert\psi_{\text{HO}}\right\vert^{2}$
and kinetic-energy density $t^{N-1}=t^{N}-\left(1/N_{\bm{k}}\right)\left(1/2\right)
\nabla\psi_{\text{HO}}^{*}\cdot\nabla\psi_{\text{HO}}$, where $\psi_{\text{HO}}$
is the orbital at the valence band maximum and is normalized to one in the unit
cell. Similarly, the contribution from the orbital $\psi_{\text{LU}}$ at the
conduction band minimum is added to $\rho^{N}$ and $t^{N}$ to calculate
$E_{\text{tot}}(N+1)$. Since in the limit of an infinite solid
($N_{\bm{k}}\rightarrow\infty$) the addition or subtraction of a single
electron has no effect on the orbitals, the three
total energies in Eq.~(\ref{eq:Eg2}) can be evaluated with the orbitals
obtained from the calculation of the neutral $N$-electron system.
Usually, DFT codes with periodic boundary conditions
deliver the total energy per unit cell (uc)
$E_{\text{tot}}^{\text{uc}}$, therefore in Eq.~(\ref{eq:Eg2})
$E_{\text{tot}}=N_{\bm{k}}E_{\text{tot}}^{\text{uc}}$.
We mention that adding or not adding a background charge to
make the $N-1$- and $N+1$-electron systems neutral leads to the same
results for $E_{\text{g}}$ when the calculation is converged with $N_{\bm{k}}$.
As in Ref.~\onlinecite{PerdewPNAS17}, we checked that Eqs.~(\ref{eq:Eg1})
and (\ref{eq:Eg2}) lead to exactly the same bandgap for a GGA.

\begin{table*}
\caption{\label{tab:band_gap1}Bandgap (in eV) of 30 solids (the space group number
is indicated in parenthesis) calculated
non-self-consistently with MGGA functionals using Eq.~(\ref{eq:Eg2}).
The GGA potential used to generate the orbitals is
EV93PW91 for MVS, mRPBE for HLE17, and RPBE for all other MGGA functionals.
The value in parenthesis is the difference
with respect to the self-consistent VASP calculation
$\left(E_{\text{g}}^{\textsc{WIEN2k}}-E_{\text{g}}^{\textsc{VASP}}\right)$.
The second set of value in parenthesis for PBE and rSCAN is the difference
with respect to \textsc{CASTEP} results.}
\begin{ruledtabular}
\begin{tabular}{lccccccccccccc} 
Solid & PBE & TPSS & revTPSS & MVS & SCAN & rSCAN & TM & HLE17 \\
\hline
Al$_{2}$O$_{3}$     (167) &   6.20 (0.01,0.00) &   6.39 (0.08) &   6.34 (0.08) &   7.30 (-0.13) &   7.20 (0.17) &   7.12 (0.06,0.07) &   6.36 (0.04) &   7.17 (0.22) \\
AlAs                (216) &   1.47 (0.04,0.01) &   1.59 (0.10) &   1.51 (0.08) &   2.33 (0.17) &   1.85 (0.12) &   1.87 (0.10,0.07) &   1.44 (0.00) &   2.76 (0.27) \\
AlN                 (186) &   4.14 (-0.00,0.01) &   4.21 (0.07) &   4.13 (0.05) &   4.89 (-0.22) &   4.78 (-0.01) &   4.82 (0.03,0.05) &   4.16 (0.00) &   4.91 (0.17) \\
AlP                 (216) &   1.59 (0.01,-0.02) &   1.74 (0.10) &   1.64 (0.08) &   2.22 (0.11) &   1.96 (0.06) &   1.95 (0.05,0.03) &   1.60 (0.02) &   2.84 (0.05) \\
AlSb                (216) &   1.22 (0.01,-0.00) &   1.31 (0.04) &   1.23 (0.04) &   1.85 (0.05) &   1.46 (0.09) &   1.46 (0.01,0.00) &   1.12 (-0.04) &   2.05 (0.25) \\
Ar                  (225) &   8.71 (-0.01,0.01) &   9.35 (0.04) &   9.26 (-0.01) &  10.37 (-0.09) &   9.58 (0.08) &   9.54 (0.01,-0.02) &   8.74 (0.07) &  10.91 (0.07) \\
BeO                 (186) &   7.37 (0.02,0.01) &   7.43 (0.10) &   7.37 (0.10) &   8.22 (-0.21) &   8.31 (0.15) &   8.25 (0.06,0.07) &   7.39 (0.05) &   8.59 (0.12) \\
BN                  (216) &   4.46 (0.01,-0.02) &   4.66 (0.23) &   4.47 (0.18) &   5.08 (0.05) &   5.06 (0.13) &   5.08 (0.09,0.08) &   4.48 (0.09) &   5.97 (0.28) \\
BP                  (216) &   1.25 (-0.02,-0.01) &   1.32 (0.01) &   1.19 (0.00) &   1.36 (-0.09) &   1.56 (-0.02) &   1.44 (-0.04,0.01) &   1.20 (-0.04) &   2.18 (-0.05) \\
C                   (227) &   4.14 (-0.01,-0.00) &   4.26 (0.07) &   4.10 (0.04) &   4.04 (-0.15) &   4.54 (-0.04) &   4.38 (-0.00,0.03) &   4.10 (0.01) &   5.14 (0.14) \\
CaF$_{2}$           (225) &   7.28 (0.00,0.02) &   7.75 (0.02) &   7.49 (-0.03) &   8.57 (0.25) &   8.04 (0.21) &   8.09 (0.22,0.22) &   6.90 (0.07) &   9.44 (0.11) \\
CaO                 (225) &   3.67 (0.04,-0.00) &   3.81 (0.06) &   3.72 (0.02) &   4.38 (0.01) &   4.45 (0.29) &   4.42 (0.26,0.23) &   3.61 (0.00) &   4.57 (0.06) \\
CdSe                (216) &   0.71 (-0.04,-0.01) &   0.94 (0.04) &   0.93 (0.04) &   2.24 (0.17) &   1.10 (0.03) &   1.18 (0.02,0.00) &   0.93 (0.03) &   1.71 (-0.02) \\
GaAs                (216) &   0.52 (-0.05,0.00) &   0.72 (0.04) &   0.76 (0.05) &   2.31 (0.15) &   0.80 (0.00) &   0.96 (-0.04,-0.03) &   0.86 (0.02) &   0.79 (0.13) \\
GaP                 (216) &   1.59 (-0.05,-0.01) &   1.70 (0.02) &   1.59 (0.02) &   2.15 (0.00) &   1.81 (-0.07) &   1.84 (-0.06,-0.00) &   1.55 (0.00) &   2.25 (0.05) \\
Ge                  (227) &   0.06 (-0.05,0.00) &   0.20 (0.03) &   0.27 (0.05) &   1.76 (0.54) &   0.24 (0.10) &   0.39 (-0.06,-0.05) &   0.42 (0.11) &   0.00 (0.00) \\
InP                 (216) &   0.68 (-0.03,-0.00) &   0.87 (0.05) &   0.85 (0.06) &   1.98 (0.07) &   0.98 (-0.07) &   1.07 (-0.04,-0.02) &   0.90 (0.04) &   1.16 (0.03) \\
KCl                 (225) &   5.21 (-0.00,0.01) &   5.70 (-0.02) &   5.59 (-0.02) &   6.61 (0.14) &   5.74 (-0.04) &   5.77 (0.00,0.00) &   5.12 (0.02) &   6.88 (-0.04) \\
Kr                  (225) &   7.26 (-0.01,-0.00) &   7.86 (-0.03) &   7.84 (-0.03) &   9.22 (0.24) &   8.00 (-0.04) &   8.04 (0.02,0.04) &   7.39 (0.04) &   9.29 (-0.02) \\
LiCl                (225) &   6.33 (0.00,0.00) &   6.54 (-0.01) &   6.56 (-0.03) &   7.75 (0.01) &   7.18 (0.00) &   7.12 (-0.02,-0.03) &   6.52 (-0.03) &   7.76 (-0.01) \\
LiF                 (225) &   9.08 (0.00,0.01) &   9.23 (-0.03) &   9.07 (-0.11) &  10.79 (0.31) &  10.14 (0.16) &  10.11 (0.14,0.11) &   8.89 (-0.07) &  10.81 (-0.01) \\
LiH                 (225) &   3.08 (0.08,0.01) &   3.37 (0.01) &   3.64 (0.02) &   3.83 (-0.19) &   3.58 (-0.06) &   3.54 (-0.03,-0.08) &   3.16 (-0.06) &   4.66 (0.04) \\
MgO                 (225) &   4.71 (0.00,0.01) &   4.80 (-0.00) &   4.73 (-0.02) &   5.88 (-0.06) &   5.69 (0.16) &   5.63 (0.07,0.08) &   4.78 (-0.01) &   5.66 (0.05) \\
NaCl                (225) &   5.11 (0.01,0.00) &   5.47 (0.02) &   5.41 (0.00) &   6.55 (-0.01) &   5.76 (-0.09) &   5.76 (-0.07,-0.06) &   5.13 (-0.05) &   6.73 (0.02) \\
NaF                 (225) &   6.33 (0.02,0.02) &   6.74 (0.06) &   6.50 (-0.02) &   7.81 (0.11) &   7.19 (0.17) &   7.15 (0.07,0.08) &   5.95 (-0.12) &   8.38 (0.09) \\
Ne                  (225) &  11.58 (-0.00,-0.01) &  12.28 (0.13) &  12.20 (-0.01) &  13.88 (0.43) &  12.97 (0.20) &  13.07 (0.17,0.07) &  11.52 (-0.07) &  14.47 (0.21) \\
Si                  (227) &   0.58 (-0.04,0.00) &   0.69 (-0.01) &   0.58 (-0.02) &   0.87 (-0.07) &   0.85 (-0.02) &   0.77 (-0.05,0.01) &   0.56 (-0.05) &   1.57 (-0.06) \\
SiC                 (216) &   1.36 (0.01,-0.03) &   1.47 (0.15) &   1.30 (0.10) &   1.82 (-0.03) &   1.78 (0.07) &   1.82 (0.08,0.07) &   1.33 (0.04) &   2.47 (0.18) \\
ZnO                 (186) &   0.82 (0.02,0.03) &   0.79 (0.06) &   0.61 (0.01) &   1.57 (0.15) &   1.32 (0.18) &   1.40 (0.17,0.19) &   0.55 (-0.08) &   2.26 (-0.12) \\
ZnS                 (216) &   2.12 (-0.03,-0.01) &   2.33 (0.04) &   2.26 (0.04) &   3.42 (0.12) &   2.60 (-0.03) &   2.66 (0.00,0.00) &   2.22 (-0.01) &   3.23 (-0.06) \\
\end{tabular}
\end{ruledtabular}
\end{table*}

\begin{table*}
\caption{\label{tab:band_gap2}Average (in eV) of the absolute
difference between the non-self-consistent (\textsc{WIEN2k}) and self-consistent
(\textsc{VASP}) bandgaps. The non-self-consistent bandgaps were calculated
with Eq.~(\ref{eq:Eg2}) for various MGGA functionals (corresponding to the
columns) by using orbitals that were generated by various potentials
(corresponding to the rows). The number in parenthesis is the average
of $\left(E_{\text{tot}}^{\text{uc}}-E_{\text{tot}}^{\text{uc},0}\right)/N_{\text{el}}^{\text{uc}}$
where $E_{\text{tot}}^{\text{uc},0}$ (in mRy) is the lowest total MGGA energy
among all those calculated using the different sets of orbitals and
$N_{\text{el}}^{\text{uc}}$ is the number of electrons per unit cell.
The results for the combination ($E_{\text{xc}}$,$v_{\text{xc}}$) that was
used for the bandgaps in Table~\ref{tab:band_gap1} are in bold.}
\begin{ruledtabular}
\begin{tabular}{lccccccccccc} 
 & TPSS & revTPSS & MVS & SCAN & rSCAN & TM & HLE17 \\
\hline
RPBE     &   $\mathbf{0.06\:(0.0)}$ &   $\mathbf{0.04\:(0.0)}$ &   0.15 (0.2) &   $\mathbf{0.10\:(0.0)}$ &   $\mathbf{0.07\:(0.0)}$ &   $\mathbf{0.04\:(0.0)}$ &   0.90 (4.9) \\
PBE      &   0.06 (0.0) &   0.05 (0.0) &   0.16 (0.5) &   0.11 (0.1) &   0.08 (0.1) &   0.04 (0.0) &   0.92 (5.3) \\
PBEsol   &   0.13 (0.2) &   0.11 (0.2) &   0.27 (0.7) &   0.17 (0.3) &   0.17 (0.3) &   0.11 (0.2) &   0.98 (6.3) \\
HCTH407  &   0.14 (0.3) &   0.15 (0.3) &   0.16 (0.3) &   0.13 (0.1) &   0.12 (0.1) &   0.18 (0.4) &   0.74 (4.0) \\
EV93PW91 &   0.10 (0.5) &   0.10 (0.4) &   $\mathbf{0.14\:(0.0)}$ &   0.08 (0.3) &   0.06 (0.3) &   0.09 (0.3) &   0.87 (4.6) \\
LDA      &   0.13 (0.7) &   0.10 (0.8) &   0.18 (1.7) &   0.18 (0.9) &   0.15 (0.9) &   0.10 (0.8) &   0.97 (7.7) \\
AK13     &   0.40 (2.6) &   0.44 (2.6) &   0.63 (1.7) &   0.49 (2.4) &   0.48 (2.3) &   0.53 (2.8) &   0.91 (4.2) \\
LB94     &   0.41 (3.7) &   0.44 (4.1) &   0.58 (5.3) &   0.41 (3.8) &   0.44 (3.8) &   0.48 (4.4) &   0.55 (4.5) \\
mRPBE    &   0.81 (4.9) &   0.83 (5.3) &   0.88 (4.8) &   0.80 (4.6) &   0.83 (4.6) &   0.89 (5.3) &   $\mathbf{0.10\:(0.0)}$ \\
HLE16    &   0.91 (6.3) &   0.92 (6.7) &   0.91 (5.8) &   0.89 (5.9) &   0.90 (5.8) &   0.99 (6.8) &   0.21 (0.5) \\
Sloc     &   1.60 (12.0) &   1.61 (12.6) &   1.58 (12.4) &   1.55 (11.6) &   1.56 (11.6) &   1.67 (12.9) &   0.78 (2.8) \\
\end{tabular}
\end{ruledtabular}
\end{table*}

\begin{figure*}
\includegraphics[width=\textwidth]{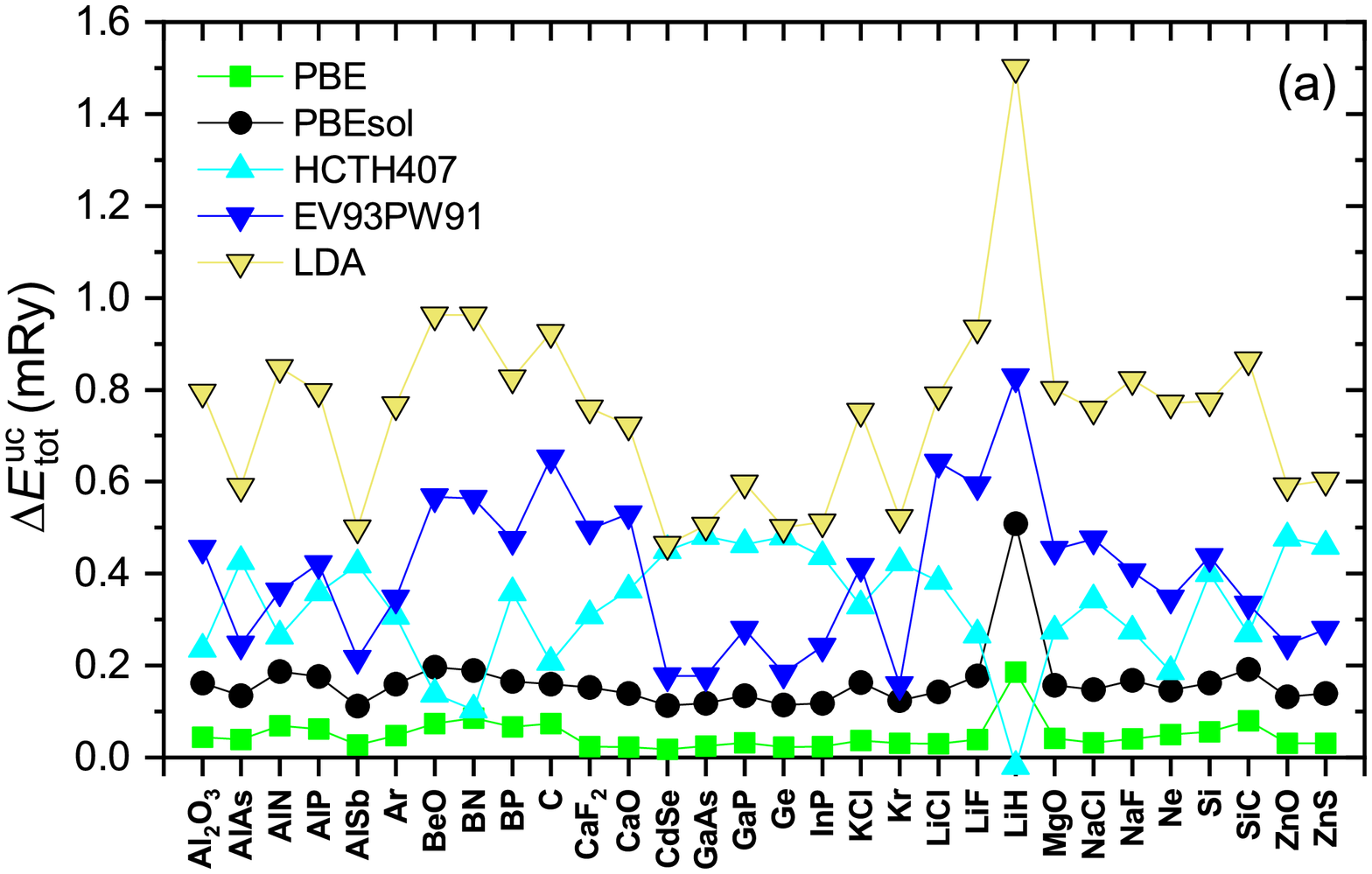}
\includegraphics[width=\textwidth]{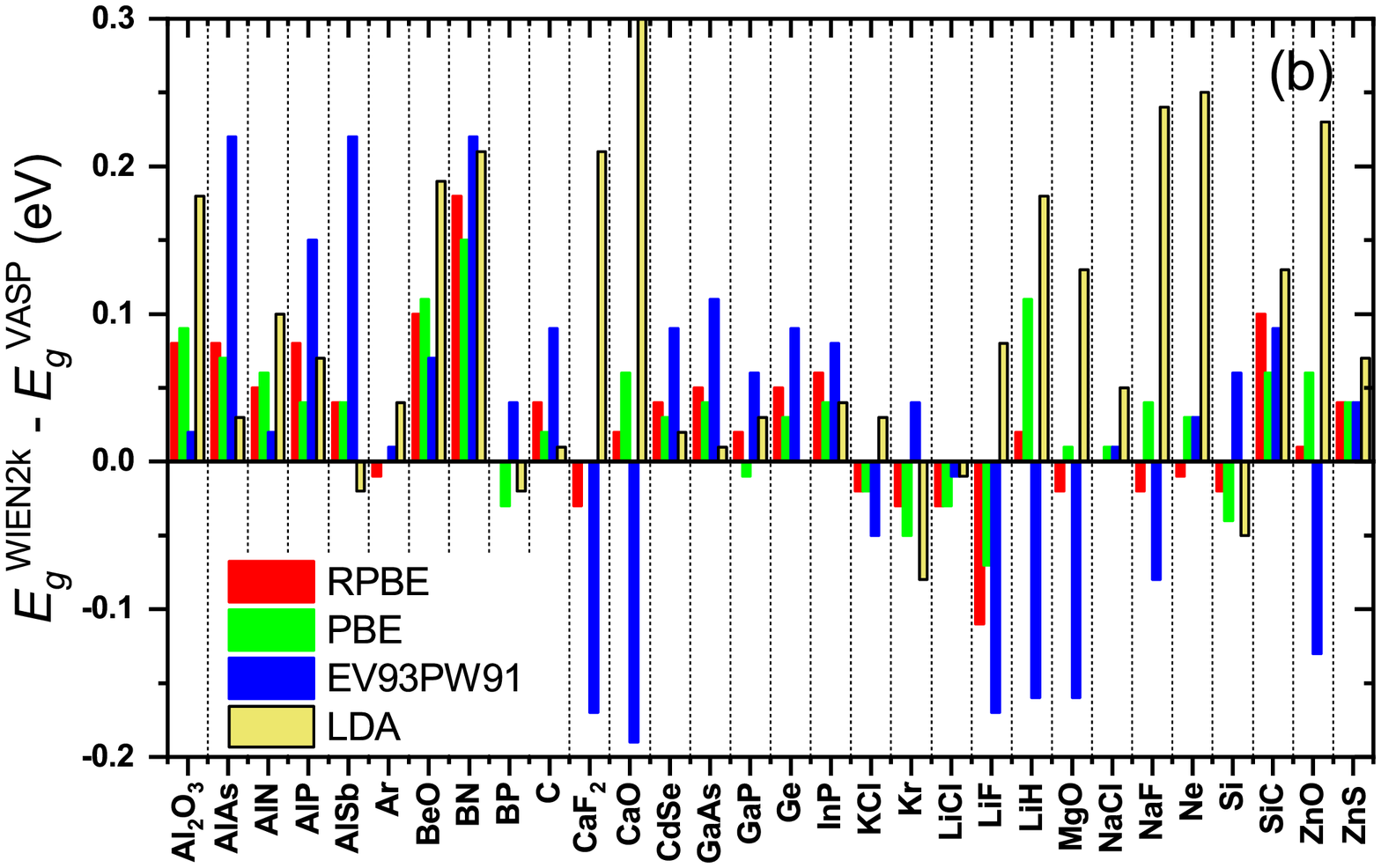}
\caption{\label{fig:revTPSS}Results obtained with the revTPSS functional using
different sets of orbitals.
(a) Total energy with respect to the value obtained with the RPBE orbitals,
$\Delta E_{\text{tot}}^{\text{uc}}=
\left(E_{\text{tot}}^{\text{uc}}[\{\psi_{i}\}]-
E_{\text{tot}}^{\text{uc}}[\{\psi_{i}^{\text{RPBE}}\}]\right)/N_{\text{el}}^{\text{uc}}$,
where $E_{\text{tot}}^{\text{uc}}$ is in mRy and
$N_{\text{el}}^{\text{uc}}$ is the number of electrons per unit cell.
(b) Bandgap difference
$E_{\text{g}}^{\textsc{WIEN2k}}-E_{\text{g}}^{\textsc{VASP}}$.
For clarity, the results with PBEsol and HCTH407 are not shown.}
\end{figure*}

With the goal of applying Eq.~(\ref{eq:Eg2}) to a MGGA non-self-consistently
without having access to the orbitals (and density) generated with the corresponding
non-multiplicative MGGA potential, the central technical question is which set of
orbitals should be used.
Our procedure is the following. For a given MGGA functional $E_{\text{xc}}$,
the total energy is evaluated with various sets of GGA orbitals.
The best set is the one leading to the lowest (i.e., most negative) total energy, since
according to the variational principle, the lower the total energy, the closer
one should approach the true MGGA orbitals of a self-consistent
calculation. Note that a similar procedure has been
used for self-interaction corrected functionals.\cite{PedersonJCP14}

In order to test the accuracy of the procedure, the MGGA energy functionals
$E_{\text{xc}}$ that we chose to calculate the bandgap are TPSS,\cite{TaoPRL03}
revTPSS,\cite{PerdewPRL09} MVS,\cite{SunPNAS15} SCAN,\cite{SunPRL15}
rSCAN,\cite{BartokJCP19} TM,\cite{TaoPRL16} and HLE17.\cite{VermaJPCC17}
The existing GGA potentials $v_{\text{xc}}$ that were used to generate the orbitals
are PBE,\cite{PerdewPRL96} RPBE,\cite{HammerPRB99}
PBEsol,\cite{PerdewPRL08} EV93PW91,\cite{EngelPRB93,PerdewPRB92b}
AK13,\cite{ArmientoPRL13} HCTH407,\cite{BoeseJCP01} and HLE16.\cite{VermaJPCL17}
An additional GGA potential that we also considered
consists of a modified RPBE potential (mRPBE), where the exchange and correlation components
are multiplied by 1.25 and 0.5, respectively. The construction of mRPBE is motivated by the
fact that the MGGA HLE17 is a modification of TPSS, with exchange and correlation
multiplied by also 1.25 and 0.5, and, as shown below, the RPBE potential is the
preferred one for generating orbitals to use with TPSS. Besides these eight GGA potentials,
the LDA,\cite{KohnPR65,PerdewPRB92a}
LB94,\cite{vanLeeuwenPRA94} and Sloc\cite{FinzelIJQC17} potentials
were also considered for generating the orbitals.
The hope is that among these multiplicative potentials,
there is one providing orbitals that are reasonably
close to the ones that would be obtained with the non-multiplicative MGGA potential.
If this is the case, then the bandgap calculated using Eq.~(\ref{eq:Eg2})
with the GGA orbitals should be close to the bandgap
$E_{\text{g}}^{\text{gKS}}=\varepsilon_{\text{LU}}-\varepsilon_{\text{HO}}$
calculated self-consistently with the MGGA potential.
Ideally, and in order to have a scheme that is useful in practice,
it should be always (or at least for most solids)
the same set of GGA orbitals (for a given MGGA)
that leads to the most negative MGGA total energy
and, hopefully, to a bandgap that is close to the true self-consistent one.

The \textsc{WIEN2k} code,\cite{WIEN2k} a full-potential and all-electron code
based on the linearized augmented plane-wave method,\cite{Singh,KarsaiCPC17}
has been used for the calculations.
For each of the 30 solids that we considered ($sp$-semiconductors,
wide bandgap ionic insulators, and rare gases, see Table~\ref{tab:band_gap1}),
the MGGA total energy was evaluated with the 11 different sets of
GGA orbitals. We found that for most solids this is the same
set that leads to the most negative MGGA total energy.
However, as expected, the set of \textit{optimal} orbitals depends on the MGGA
functional under consideration. Those are the ones that were generated from
EV93PW91 for MVS, mRPBE for HLE17, and RPBE for the other MGGAs.

Using these \textit{optimal} orbitals, the results for the bandgap are shown in
Table~\ref{tab:band_gap1}.
The calculations were done at the experimental geometry.\cite{BorlidoJCTC19}
Comparison is made with the bandgaps obtained using the
\textsc{VASP} code\cite{KressePRB96} (based on the projector augmented wave
method\cite{BlochlPRB94b}), which allows for self-consistent MGGA calculations
(details of the calculations can be found in Ref.~\onlinecite{BorlidoJCTC19}).
Additional self-consistent pseudopotential calculations for the PBE and
rSCAN functionals (with pseudopotentials generated specifically for the
respective functional) were
performed with the \textsc{CASTEP} code.\cite{PayneRMP92,ClarkZKCM05}
It is shown that the agreement between \textsc{WIEN2k} and \textsc{VASP} is often
very good, since the difference is below 0.1~eV in the majority of cases except for
MVS (using the EV93PW91 orbitals).
The largest discrepancies, 0.54~eV for Ge and 0.43~eV for Ne,
were obtained with MVS. In the case of Ne, a discrepancy of this
order of magnitude is acceptable since it is rather small compared to the bandgap
which is above 13~eV. However, this is not the case for Ge, since
0.54~eV represents 45\% of the MVS bandgap of 1.22~eV calculated with \textsc{VASP}.
Other differences between \textsc{WIEN2k} and \textsc{VASP} which are relatively
important are obtained for AlAs and AlSb with HLE17, and
for ZnO with SCAN and rSCAN. Actually, for the latter functional the \textsc{CASTEP}
bandgaps agree in general extremely well with those from \textsc{VASP}
(in the same way as with PBE), which indicates that the
\textsc{WIEN2k}/\textsc{VASP} discrepancies should be due to
the non-self-consistent procedure.
MVS and HLE17 lead to bandgaps that are clearly larger than with all other functionals,
therefore these two functionals are somehow different. This means that
for these two functionals, some (occasional) non-negligible error due to the use
of an inconsistent pseudopotential should not be completely excluded.

The average over all solids of the absolute difference
$\left\vert E_{\text{g}}^{\textsc{WIEN2k}}-E_{\text{g}}^{\textsc{VASP}}\right\vert$
is shown in Table~\ref{tab:band_gap2} for all combinations
($E_{\text{xc}}$,$v_{\text{xc}}$), i.e., all sets of orbitals (generated by the
various $v_{\text{xc}}$) plugged into all MGGA energy functionals $E_{\text{xc}}$.
The values with the optimal orbitals (highlighted in bold) represent
the statistics of the results in Table~\ref{tab:band_gap1}. For all
functionals except MVS the average error is below 0.1~eV.
A larger value of 0.14~eV is obtained for the MVS functional with the EV93PW91 orbitals.
We checked that combining the EV93 exchange potential\cite{EngelPRB93} with other
correlation potentials like LDA,\cite{PerdewPRB92a} PBE,\cite{PerdewPRL96}
or LYP\cite{LeePRB88} does not improve the MVS results.

In general, a clear correlation between the total MGGA energy and the difference
in the bandgap can be observed; if a set of orbitals leads to
(one of) the most negative total energy for a particular MGGA functional
(quantified by the average of the total energy per cell and per electron,
see Table~\ref{tab:band_gap2}), then the agreement with \textsc{VASP} for
the bandgap will be one of the best. However, choosing a set
of orbitals that is not the one that minimizes the MGGA functional may
seriously degrade the agreement with \textsc{VASP}.
For instance, the mRPBE orbitals, which are the optimal for the
MGGA HLE17, lead to rather inaccurate results for all other MGGAs
(the disagreement with \textsc{VASP} is $0.8$-$0.9$~eV).
Among all sets of orbitals that we have
considered, the Sloc orbitals lead most of the time to the least
negative MGGA total energy and, consequently, to the worst agreement
with \textsc{VASP} results for the bandgap except with HLE17.
We just note that in the case of the (r)SCAN functional, the EV93PW91
orbitals, despite being less optimal than the RPBE ones for the total energy,
lead to slightly better agreement between \textsc{WIEN2k} and \textsc{VASP}
for the bandgap. However, the differences are at the level of 0.01-0.02~eV,
which is very small and of the same order as errors that could come
from other parameters like the basis set size or the pseudopotential.
A graphical illustration of the detailed results is shown
in Fig.~\ref{fig:revTPSS} for the revTPSS functional when it
is evaluated using the orbitals obtained from some of the best potentials.
From Fig.~\ref{fig:revTPSS}(a), which compares the revTPSS total energies,
we can see that the ordering of the potentials, except HCTH407, is the
same for all solids. The results with the HCTH407 orbitals
alternate with those from the other sets of orbitals.
The results for the bandgap in Fig.~\ref{fig:revTPSS}(b)
show that using the RPBE orbitals (the optimal ones for the revTPSS total energy)
does not systematically lead to the smallest difference
$E_{\text{g}}^{\textsc{WIEN2k}}-E_{\text{g}}^{\textsc{VASP}}$,
however on average the difference is the smallest (0.04~eV, see Table~\ref{tab:band_gap2}).
The results for the revTPSS bandgap with
the PBEsol and HCTH407 orbitals [not shown in Fig.~\ref{fig:revTPSS}(b)] exhibit
for a few cases (e,g., Ge with PBEsol) rather large errors.

We mention that Lima \textit{et al}.\cite{LimaJCP07} proposed to approximate
the potential $v_{\text{xc}}^{\text{B}}$ of a functional
$E_{\text{xc}}^{\text{B}}=\int\epsilon_{\text{xc}}^{\text{B}}d^{3}r$ as the
rescaling of the potential of another functional A:
$v_{\text{xc}}^{\text{B}}\approx
\left(\epsilon_{\text{xc}}^{\text{B}}/\epsilon_{\text{xc}}^{\text{A}}\right)
v_{\text{xc}}^{\text{A}}$. This method may be useful when the potential
$v_{\text{xc}}^{\text{B}}$ is not implemented, as in our case here with MGGAs.
We tested this scheme for a few cases, including
$\text{A}=\text{PBE}$, RPBE, or EV93PW91 and $\text{B}=\text{SCAN}$ or MVS,
to obtain an approximate (and multiplicative) MGGA potential that is used
to calculate the orbitals that are then plugged into the corresponding
total-energy MGGA functional. However, the results (not shown) are typically worse
than those obtained with several of the GGA potentials, meaning that
the orbitals obtained with $v_{\text{xc}}^\text{B=MGGA}$ are
not particularly close to the true MGGA orbitals.
Thus the method does not seem to be really useful for our purpose.

In summary, we have shown that the use of total energies [Eq.~(\ref{eq:Eg2})]
to calculate the bandgap of MGGA functionals can lead to very accurate results even
when GGA orbitals are used. However, it is important to choose reasonable
orbitals, i.e., orbitals that satisfy the variational principle as much
as possible. Luckily, once a GGA potential to generate the orbitals has
been shown to be appropriate for a MGGA functional in a few cases,
then it appears to be rather safe to use it for other solids. Thus, this scheme
allows to obtain the bandgap with MGGA functionals when the corresponding
potential is not implemented. In principle, this simple procedure can be
applied to any kind of (new) energy functionals, whose self-consistent
implementation would require intense effort.

J.D. and P.B. acknowledge support from the Austrian Science Fund
(FWF) through project W1243 (Solids4Fun).
S.B. acknowledges partial support from the DFG through the project
BO 4280/8-1. Computational resources were provided by the
Leibniz Supercomputing Centre through the projects pr62ja.
M.A.L.M. acknowledges partial support from the German DFG through
the project MA6787/6-1.
A.P.B. acknowledges support from the Collaborative Computational Project for
NMR Crystallography (CCP-NC) and UKCP Consortium, both funded by the
Engineering and Physical Sciences Research Council (EPSRC) under Grants
No. EP/M022501/1 and No. EP/P022561/1, respectively. Some of the calculations
were run using the STFC Scientific Computing Department's SCARF cluster.

\bibliography{references}

\begin{thebibliography}{62}%
\makeatletter
\providecommand \@ifxundefined [1]{%
 \@ifx{#1\undefined}
}%
\providecommand \@ifnum [1]{%
 \ifnum #1\expandafter \@firstoftwo
 \else \expandafter \@secondoftwo
 \fi
}%
\providecommand \@ifx [1]{%
 \ifx #1\expandafter \@firstoftwo
 \else \expandafter \@secondoftwo
 \fi
}%
\providecommand \natexlab [1]{#1}%
\providecommand \enquote  [1]{``#1''}%
\providecommand \bibnamefont  [1]{#1}%
\providecommand \bibfnamefont [1]{#1}%
\providecommand \citenamefont [1]{#1}%
\providecommand \href@noop [0]{\@secondoftwo}%
\providecommand \href [0]{\begingroup \@sanitize@url \@href}%
\providecommand \@href[1]{\@@startlink{#1}\@@href}%
\providecommand \@@href[1]{\endgroup#1\@@endlink}%
\providecommand \@sanitize@url [0]{\catcode `\\12\catcode `\$12\catcode
  `\&12\catcode `\#12\catcode `\^12\catcode `\_12\catcode `\%12\relax}%
\providecommand \@@startlink[1]{}%
\providecommand \@@endlink[0]{}%
\providecommand \url  [0]{\begingroup\@sanitize@url \@url }%
\providecommand \@url [1]{\endgroup\@href {#1}{\urlprefix }}%
\providecommand \urlprefix  [0]{URL }%
\providecommand \Eprint [0]{\href }%
\providecommand \doibase [0]{http://dx.doi.org/}%
\providecommand \selectlanguage [0]{\@gobble}%
\providecommand \bibinfo  [0]{\@secondoftwo}%
\providecommand \bibfield  [0]{\@secondoftwo}%
\providecommand \translation [1]{[#1]}%
\providecommand \BibitemOpen [0]{}%
\providecommand \bibitemStop [0]{}%
\providecommand \bibitemNoStop [0]{.\EOS\space}%
\providecommand \EOS [0]{\spacefactor3000\relax}%
\providecommand \BibitemShut  [1]{\csname bibitem#1\endcsname}%
\let\auto@bib@innerbib\@empty
%</preamble>
\bibitem [{\citenamefont {Hohenberg}\ and\ \citenamefont
  {Kohn}(1964)}]{HohenbergPR64}%
  \BibitemOpen
  \bibfield  {author} {\bibinfo {author} {\bibfnamefont {P.}~\bibnamefont
  {Hohenberg}}\ and\ \bibinfo {author} {\bibfnamefont {W.}~\bibnamefont
  {Kohn}},\ }\href@noop {} {\bibfield  {journal} {\bibinfo  {journal} {Phys.
  Rev.}\ }\textbf {\bibinfo {volume} {136}},\ \bibinfo {pages} {B864} (\bibinfo
  {year} {1964})}\BibitemShut {NoStop}%
\bibitem [{\citenamefont {Kohn}\ and\ \citenamefont {Sham}(1965)}]{KohnPR65}%
  \BibitemOpen
  \bibfield  {author} {\bibinfo {author} {\bibfnamefont {W.}~\bibnamefont
  {Kohn}}\ and\ \bibinfo {author} {\bibfnamefont {L.~J.}\ \bibnamefont
  {Sham}},\ }\href@noop {} {\bibfield  {journal} {\bibinfo  {journal} {Phys.
  Rev.}\ }\textbf {\bibinfo {volume} {140}},\ \bibinfo {pages} {A1133}
  (\bibinfo {year} {1965})}\BibitemShut {NoStop}%
\bibitem [{\citenamefont {Seidl}\ \emph {et~al.}(1996)\citenamefont {Seidl},
  \citenamefont {G\"orling}, \citenamefont {Vogl}, \citenamefont {Majewski},\
  and\ \citenamefont {Levy}}]{SeidlPRB96}%
  \BibitemOpen
  \bibfield  {author} {\bibinfo {author} {\bibfnamefont {A.}~\bibnamefont
  {Seidl}}, \bibinfo {author} {\bibfnamefont {A.}~\bibnamefont {G\"orling}},
  \bibinfo {author} {\bibfnamefont {P.}~\bibnamefont {Vogl}}, \bibinfo {author}
  {\bibfnamefont {J.~A.}\ \bibnamefont {Majewski}}, \ and\ \bibinfo {author}
  {\bibfnamefont {M.}~\bibnamefont {Levy}},\ }\href@noop {} {\bibfield
  {journal} {\bibinfo  {journal} {Phys. Rev. B}\ }\textbf {\bibinfo {volume}
  {53}},\ \bibinfo {pages} {3764} (\bibinfo {year} {1996})}\BibitemShut
  {NoStop}%
\bibitem [{\citenamefont {Perdew}\ \emph {et~al.}(1982)\citenamefont {Perdew},
  \citenamefont {Parr}, \citenamefont {Levy},\ and\ \citenamefont
  {Balduz}}]{PerdewPRL82}%
  \BibitemOpen
  \bibfield  {author} {\bibinfo {author} {\bibfnamefont {J.~P.}\ \bibnamefont
  {Perdew}}, \bibinfo {author} {\bibfnamefont {R.~G.}\ \bibnamefont {Parr}},
  \bibinfo {author} {\bibfnamefont {M.}~\bibnamefont {Levy}}, \ and\ \bibinfo
  {author} {\bibfnamefont {J.~L.}\ \bibnamefont {Balduz}, \bibfnamefont
  {Jr.}},\ }\href@noop {} {\bibfield  {journal} {\bibinfo  {journal} {Phys.
  Rev. Lett.}\ }\textbf {\bibinfo {volume} {49}},\ \bibinfo {pages} {1691}
  (\bibinfo {year} {1982})}\BibitemShut {NoStop}%
\bibitem [{\citenamefont {Sham}\ and\ \citenamefont
  {Schl\"uter}(1983)}]{ShamPRL83}%
  \BibitemOpen
  \bibfield  {author} {\bibinfo {author} {\bibfnamefont {L.~J.}\ \bibnamefont
  {Sham}}\ and\ \bibinfo {author} {\bibfnamefont {M.}~\bibnamefont
  {Schl\"uter}},\ }\href@noop {} {\bibfield  {journal} {\bibinfo  {journal}
  {Phys. Rev. Lett.}\ }\textbf {\bibinfo {volume} {51}},\ \bibinfo {pages}
  {1888} (\bibinfo {year} {1983})}\BibitemShut {NoStop}%
\bibitem [{\citenamefont {Perdew}, \citenamefont {Burke},\ and\ \citenamefont
  {Ernzerhof}(1996)}]{PerdewPRL96}%
  \BibitemOpen
  \bibfield  {author} {\bibinfo {author} {\bibfnamefont {J.~P.}\ \bibnamefont
  {Perdew}}, \bibinfo {author} {\bibfnamefont {K.}~\bibnamefont {Burke}}, \
  and\ \bibinfo {author} {\bibfnamefont {M.}~\bibnamefont {Ernzerhof}},\
  }\href@noop {} {\bibfield  {journal} {\bibinfo  {journal} {Phys. Rev. Lett.}\
  }\textbf {\bibinfo {volume} {77}},\ \bibinfo {pages} {3865} (\bibinfo {year}
  {1996})},\ \bibinfo {note} {\textbf{78}, 1396(E) (1997)}\BibitemShut
  {NoStop}%
\bibitem [{\citenamefont {Becke}(1988)}]{BeckePRA88}%
  \BibitemOpen
  \bibfield  {author} {\bibinfo {author} {\bibfnamefont {A.~D.}\ \bibnamefont
  {Becke}},\ }\href@noop {} {\bibfield  {journal} {\bibinfo  {journal} {Phys.
  Rev. A}\ }\textbf {\bibinfo {volume} {38}},\ \bibinfo {pages} {3098}
  (\bibinfo {year} {1988})}\BibitemShut {NoStop}%
\bibitem [{\citenamefont {Lee}, \citenamefont {Yang},\ and\ \citenamefont
  {Parr}(1988)}]{LeePRB88}%
  \BibitemOpen
  \bibfield  {author} {\bibinfo {author} {\bibfnamefont {C.}~\bibnamefont
  {Lee}}, \bibinfo {author} {\bibfnamefont {W.}~\bibnamefont {Yang}}, \ and\
  \bibinfo {author} {\bibfnamefont {R.~G.}\ \bibnamefont {Parr}},\ }\href@noop
  {} {\bibfield  {journal} {\bibinfo  {journal} {Phys. Rev. B}\ }\textbf
  {\bibinfo {volume} {37}},\ \bibinfo {pages} {785} (\bibinfo {year}
  {1988})}\BibitemShut {NoStop}%
\bibitem [{\citenamefont {Heyd}\ \emph {et~al.}(2005)\citenamefont {Heyd},
  \citenamefont {Peralta}, \citenamefont {Scuseria},\ and\ \citenamefont
  {Martin}}]{HeydJCP05}%
  \BibitemOpen
  \bibfield  {author} {\bibinfo {author} {\bibfnamefont {J.}~\bibnamefont
  {Heyd}}, \bibinfo {author} {\bibfnamefont {J.~E.}\ \bibnamefont {Peralta}},
  \bibinfo {author} {\bibfnamefont {G.~E.}\ \bibnamefont {Scuseria}}, \ and\
  \bibinfo {author} {\bibfnamefont {R.~L.}\ \bibnamefont {Martin}},\
  }\href@noop {} {\bibfield  {journal} {\bibinfo  {journal} {J. Chem. Phys.}\
  }\textbf {\bibinfo {volume} {123}},\ \bibinfo {pages} {174101} (\bibinfo
  {year} {2005})}\BibitemShut {NoStop}%
\bibitem [{\citenamefont {Kraisler}\ and\ \citenamefont
  {Kronik}(2014)}]{KraislerJCP14}%
  \BibitemOpen
  \bibfield  {author} {\bibinfo {author} {\bibfnamefont {E.}~\bibnamefont
  {Kraisler}}\ and\ \bibinfo {author} {\bibfnamefont {L.}~\bibnamefont
  {Kronik}},\ }\href@noop {} {\bibfield  {journal} {\bibinfo  {journal} {J.
  Chem. Phys.}\ }\textbf {\bibinfo {volume} {140}},\ \bibinfo {pages} {18A540}
  (\bibinfo {year} {2014})}\BibitemShut {NoStop}%
\bibitem [{\citenamefont {G\"orling}(2015)}]{GorlingPRB15}%
  \BibitemOpen
  \bibfield  {author} {\bibinfo {author} {\bibfnamefont {A.}~\bibnamefont
  {G\"orling}},\ }\href@noop {} {\bibfield  {journal} {\bibinfo  {journal}
  {Phys. Rev. B}\ }\textbf {\bibinfo {volume} {91}},\ \bibinfo {pages} {245120}
  (\bibinfo {year} {2015})}\BibitemShut {NoStop}%
\bibitem [{\citenamefont {Perdew}\ and\ \citenamefont
  {Zunger}(1981)}]{PerdewPRB81}%
  \BibitemOpen
  \bibfield  {author} {\bibinfo {author} {\bibfnamefont {J.~P.}\ \bibnamefont
  {Perdew}}\ and\ \bibinfo {author} {\bibfnamefont {A.}~\bibnamefont
  {Zunger}},\ }\href@noop {} {\bibfield  {journal} {\bibinfo  {journal} {Phys.
  Rev. B}\ }\textbf {\bibinfo {volume} {23}},\ \bibinfo {pages} {5048}
  (\bibinfo {year} {1981})}\BibitemShut {NoStop}%
\bibitem [{\citenamefont {Mori-S\'{a}nchez}, \citenamefont {Cohen},\ and\
  \citenamefont {Yang}(2008)}]{MoriSanchezPRL08}%
  \BibitemOpen
  \bibfield  {author} {\bibinfo {author} {\bibfnamefont {P.}~\bibnamefont
  {Mori-S\'{a}nchez}}, \bibinfo {author} {\bibfnamefont {A.~J.}\ \bibnamefont
  {Cohen}}, \ and\ \bibinfo {author} {\bibfnamefont {W.}~\bibnamefont {Yang}},\
  }\href@noop {} {\bibfield  {journal} {\bibinfo  {journal} {Phys. Rev. Lett.}\
  }\textbf {\bibinfo {volume} {100}},\ \bibinfo {pages} {146401} (\bibinfo
  {year} {2008})}\BibitemShut {NoStop}%
\bibitem [{\citenamefont {Perdew}(1990)}]{PerdewAQC90}%
  \BibitemOpen
  \bibfield  {author} {\bibinfo {author} {\bibfnamefont {J.~P.}\ \bibnamefont
  {Perdew}},\ }\href@noop {} {\bibfield  {journal} {\bibinfo  {journal} {Adv.
  Quantum Chem.}\ }\textbf {\bibinfo {volume} {21}},\ \bibinfo {pages} {113}
  (\bibinfo {year} {1990})}\BibitemShut {NoStop}%
\bibitem [{\citenamefont {Tran}\ and\ \citenamefont {Blaha}(2009)}]{TranPRL09}%
  \BibitemOpen
  \bibfield  {author} {\bibinfo {author} {\bibfnamefont {F.}~\bibnamefont
  {Tran}}\ and\ \bibinfo {author} {\bibfnamefont {P.}~\bibnamefont {Blaha}},\
  }\href@noop {} {\bibfield  {journal} {\bibinfo  {journal} {Phys. Rev. Lett.}\
  }\textbf {\bibinfo {volume} {102}},\ \bibinfo {pages} {226401} (\bibinfo
  {year} {2009})}\BibitemShut {NoStop}%
\bibitem [{\citenamefont {Armiento}\ and\ \citenamefont
  {K\"{u}mmel}(2013)}]{ArmientoPRL13}%
  \BibitemOpen
  \bibfield  {author} {\bibinfo {author} {\bibfnamefont {R.}~\bibnamefont
  {Armiento}}\ and\ \bibinfo {author} {\bibfnamefont {S.}~\bibnamefont
  {K\"{u}mmel}},\ }\href@noop {} {\bibfield  {journal} {\bibinfo  {journal}
  {Phys. Rev. Lett.}\ }\textbf {\bibinfo {volume} {111}},\ \bibinfo {pages}
  {036402} (\bibinfo {year} {2013})}\BibitemShut {NoStop}%
\bibitem [{\citenamefont {Verma}\ and\ \citenamefont
  {Truhlar}(2017{\natexlab{a}})}]{VermaJPCL17}%
  \BibitemOpen
  \bibfield  {author} {\bibinfo {author} {\bibfnamefont {P.}~\bibnamefont
  {Verma}}\ and\ \bibinfo {author} {\bibfnamefont {D.~G.}\ \bibnamefont
  {Truhlar}},\ }\href@noop {} {\bibfield  {journal} {\bibinfo  {journal} {J.
  Phys. Chem. Lett.}\ }\textbf {\bibinfo {volume} {8}},\ \bibinfo {pages} {380}
  (\bibinfo {year} {2017}{\natexlab{a}})}\BibitemShut {NoStop}%
\bibitem [{\citenamefont {Della~Sala}, \citenamefont {Fabiano},\ and\
  \citenamefont {Constantin}(2016)}]{DellaSalaIJQC16}%
  \BibitemOpen
  \bibfield  {author} {\bibinfo {author} {\bibfnamefont {F.}~\bibnamefont
  {Della~Sala}}, \bibinfo {author} {\bibfnamefont {E.}~\bibnamefont {Fabiano}},
  \ and\ \bibinfo {author} {\bibfnamefont {L.~A.}\ \bibnamefont {Constantin}},\
  }\href@noop {} {\bibfield  {journal} {\bibinfo  {journal} {Int. J. Quantum
  Chem.}\ }\textbf {\bibinfo {volume} {116}},\ \bibinfo {pages} {1641}
  (\bibinfo {year} {2016})}\BibitemShut {NoStop}%
\bibitem [{\citenamefont {Becke}(1993)}]{BeckeJCP93b}%
  \BibitemOpen
  \bibfield  {author} {\bibinfo {author} {\bibfnamefont {A.~D.}\ \bibnamefont
  {Becke}},\ }\href@noop {} {\bibfield  {journal} {\bibinfo  {journal} {J.
  Chem. Phys.}\ }\textbf {\bibinfo {volume} {98}},\ \bibinfo {pages} {5648}
  (\bibinfo {year} {1993})}\BibitemShut {NoStop}%
\bibitem [{\citenamefont {K{\"u}mmel}\ and\ \citenamefont
  {Kronik}(2008)}]{KuemmelRMP08}%
  \BibitemOpen
  \bibfield  {author} {\bibinfo {author} {\bibfnamefont {S.}~\bibnamefont
  {K{\"u}mmel}}\ and\ \bibinfo {author} {\bibfnamefont {L.}~\bibnamefont
  {Kronik}},\ }\href@noop {} {\bibfield  {journal} {\bibinfo  {journal} {Rev.
  Mod. Phys.}\ }\textbf {\bibinfo {volume} {80}},\ \bibinfo {pages} {3}
  (\bibinfo {year} {2008})}\BibitemShut {NoStop}%
\bibitem [{\citenamefont {Yang}, \citenamefont {Cohen},\ and\ \citenamefont
  {Mori-S\'{a}nchez}(2012)}]{YangJCP12}%
  \BibitemOpen
  \bibfield  {author} {\bibinfo {author} {\bibfnamefont {W.}~\bibnamefont
  {Yang}}, \bibinfo {author} {\bibfnamefont {A.~J.}\ \bibnamefont {Cohen}}, \
  and\ \bibinfo {author} {\bibfnamefont {P.}~\bibnamefont {Mori-S\'{a}nchez}},\
  }\href@noop {} {\bibfield  {journal} {\bibinfo  {journal} {J. Chem. Phys.}\
  }\textbf {\bibinfo {volume} {136}},\ \bibinfo {pages} {204111} (\bibinfo
  {year} {2012})}\BibitemShut {NoStop}%
\bibitem [{\citenamefont {Yang}\ \emph {et~al.}(2016)\citenamefont {Yang},
  \citenamefont {Peng}, \citenamefont {Sun},\ and\ \citenamefont
  {Perdew}}]{YangPRB16}%
  \BibitemOpen
  \bibfield  {author} {\bibinfo {author} {\bibfnamefont {Z.-h.}\ \bibnamefont
  {Yang}}, \bibinfo {author} {\bibfnamefont {H.}~\bibnamefont {Peng}}, \bibinfo
  {author} {\bibfnamefont {J.}~\bibnamefont {Sun}}, \ and\ \bibinfo {author}
  {\bibfnamefont {J.~P.}\ \bibnamefont {Perdew}},\ }\href@noop {} {\bibfield
  {journal} {\bibinfo  {journal} {Phys. Rev. B}\ }\textbf {\bibinfo {volume}
  {93}},\ \bibinfo {pages} {205205} (\bibinfo {year} {2016})}\BibitemShut
  {NoStop}%
\bibitem [{\citenamefont {Perdew}\ \emph {et~al.}(2017)\citenamefont {Perdew},
  \citenamefont {Yang}, \citenamefont {Burke}, \citenamefont {Yang},
  \citenamefont {Gross}, \citenamefont {Scheffler}, \citenamefont {Scuseria},
  \citenamefont {Henderson}, \citenamefont {Zhang}, \citenamefont {Ruzsinszky},
  \citenamefont {Peng}, \citenamefont {Sun}, \citenamefont {Trushin},\ and\
  \citenamefont {G\"{o}rling}}]{PerdewPNAS17}%
  \BibitemOpen
  \bibfield  {author} {\bibinfo {author} {\bibfnamefont {J.~P.}\ \bibnamefont
  {Perdew}}, \bibinfo {author} {\bibfnamefont {W.}~\bibnamefont {Yang}},
  \bibinfo {author} {\bibfnamefont {K.}~\bibnamefont {Burke}}, \bibinfo
  {author} {\bibfnamefont {Z.}~\bibnamefont {Yang}}, \bibinfo {author}
  {\bibfnamefont {E.~K.~U.}\ \bibnamefont {Gross}}, \bibinfo {author}
  {\bibfnamefont {M.}~\bibnamefont {Scheffler}}, \bibinfo {author}
  {\bibfnamefont {G.~E.}\ \bibnamefont {Scuseria}}, \bibinfo {author}
  {\bibfnamefont {T.~M.}\ \bibnamefont {Henderson}}, \bibinfo {author}
  {\bibfnamefont {I.~Y.}\ \bibnamefont {Zhang}}, \bibinfo {author}
  {\bibfnamefont {A.}~\bibnamefont {Ruzsinszky}}, \bibinfo {author}
  {\bibfnamefont {H.}~\bibnamefont {Peng}}, \bibinfo {author} {\bibfnamefont
  {J.}~\bibnamefont {Sun}}, \bibinfo {author} {\bibfnamefont {E.}~\bibnamefont
  {Trushin}}, \ and\ \bibinfo {author} {\bibfnamefont {A.}~\bibnamefont
  {G\"{o}rling}},\ }\href@noop {} {\bibfield  {journal} {\bibinfo  {journal}
  {Proc. Natl. Acad. Sci. U.S.A.}\ }\textbf {\bibinfo {volume} {114}},\
  \bibinfo {pages} {2801} (\bibinfo {year} {2017})}\BibitemShut {NoStop}%
\bibitem [{\citenamefont {Sun}, \citenamefont {Ruzsinszky},\ and\ \citenamefont
  {Perdew}(2015)}]{SunPRL15}%
  \BibitemOpen
  \bibfield  {author} {\bibinfo {author} {\bibfnamefont {J.}~\bibnamefont
  {Sun}}, \bibinfo {author} {\bibfnamefont {A.}~\bibnamefont {Ruzsinszky}}, \
  and\ \bibinfo {author} {\bibfnamefont {J.~P.}\ \bibnamefont {Perdew}},\
  }\href@noop {} {\bibfield  {journal} {\bibinfo  {journal} {Phys. Rev. Lett.}\
  }\textbf {\bibinfo {volume} {115}},\ \bibinfo {pages} {036402} (\bibinfo
  {year} {2015})}\BibitemShut {NoStop}%
\bibitem [{\citenamefont {Isaacs}\ and\ \citenamefont
  {Wolverton}(2018)}]{IsaacsPRM18}%
  \BibitemOpen
  \bibfield  {author} {\bibinfo {author} {\bibfnamefont {E.~B.}\ \bibnamefont
  {Isaacs}}\ and\ \bibinfo {author} {\bibfnamefont {C.}~\bibnamefont
  {Wolverton}},\ }\href@noop {} {\bibfield  {journal} {\bibinfo  {journal}
  {Phys. Rev. Materials}\ }\textbf {\bibinfo {volume} {2}},\ \bibinfo {pages}
  {063801} (\bibinfo {year} {2018})}\BibitemShut {NoStop}%
\bibitem [{\citenamefont {Zhang}\ \emph {et~al.}(2018)\citenamefont {Zhang},
  \citenamefont {Kitchaev}, \citenamefont {Yang}, \citenamefont {Chen},
  \citenamefont {Dacek}, \citenamefont {Sarmiento-P\'{e}rez}, \citenamefont
  {Marques}, \citenamefont {Peng}, \citenamefont {Ceder}, \citenamefont
  {Perdew},\ and\ \citenamefont {Sun}}]{ZhangNPJCM18}%
  \BibitemOpen
  \bibfield  {author} {\bibinfo {author} {\bibfnamefont {Y.}~\bibnamefont
  {Zhang}}, \bibinfo {author} {\bibfnamefont {D.~A.}\ \bibnamefont {Kitchaev}},
  \bibinfo {author} {\bibfnamefont {J.}~\bibnamefont {Yang}}, \bibinfo {author}
  {\bibfnamefont {T.}~\bibnamefont {Chen}}, \bibinfo {author} {\bibfnamefont
  {S.~T.}\ \bibnamefont {Dacek}}, \bibinfo {author} {\bibfnamefont {R.~A.}\
  \bibnamefont {Sarmiento-P\'{e}rez}}, \bibinfo {author} {\bibfnamefont
  {M.~A.~L.}\ \bibnamefont {Marques}}, \bibinfo {author} {\bibfnamefont
  {H.}~\bibnamefont {Peng}}, \bibinfo {author} {\bibfnamefont {G.}~\bibnamefont
  {Ceder}}, \bibinfo {author} {\bibfnamefont {J.~P.}\ \bibnamefont {Perdew}}, \
  and\ \bibinfo {author} {\bibfnamefont {J.}~\bibnamefont {Sun}},\ }\href@noop
  {} {\bibfield  {journal} {\bibinfo  {journal} {npj Comput. Mater.}\ }\textbf
  {\bibinfo {volume} {4}},\ \bibinfo {pages} {9} (\bibinfo {year}
  {2018})}\BibitemShut {NoStop}%
\bibitem [{\citenamefont {Lane}\ \emph {et~al.}(2018)\citenamefont {Lane},
  \citenamefont {Furness}, \citenamefont {Buda}, \citenamefont {Zhang},
  \citenamefont {Markiewicz}, \citenamefont {Barbiellini}, \citenamefont
  {Sun},\ and\ \citenamefont {Bansil}}]{LanePRB18}%
  \BibitemOpen
  \bibfield  {author} {\bibinfo {author} {\bibfnamefont {C.}~\bibnamefont
  {Lane}}, \bibinfo {author} {\bibfnamefont {J.~W.}\ \bibnamefont {Furness}},
  \bibinfo {author} {\bibfnamefont {I.~G.}\ \bibnamefont {Buda}}, \bibinfo
  {author} {\bibfnamefont {Y.}~\bibnamefont {Zhang}}, \bibinfo {author}
  {\bibfnamefont {R.~S.}\ \bibnamefont {Markiewicz}}, \bibinfo {author}
  {\bibfnamefont {B.}~\bibnamefont {Barbiellini}}, \bibinfo {author}
  {\bibfnamefont {J.}~\bibnamefont {Sun}}, \ and\ \bibinfo {author}
  {\bibfnamefont {A.}~\bibnamefont {Bansil}},\ }\href@noop {} {\bibfield
  {journal} {\bibinfo  {journal} {Phys. Rev. B}\ }\textbf {\bibinfo {volume}
  {98}},\ \bibinfo {pages} {125140} (\bibinfo {year} {2018})}\BibitemShut
  {NoStop}%
\bibitem [{\citenamefont {Varignon}, \citenamefont {Bibes},\ and\ \citenamefont
  {Zunger}(2019)}]{VarignonPRB19}%
  \BibitemOpen
  \bibfield  {author} {\bibinfo {author} {\bibfnamefont {J.}~\bibnamefont
  {Varignon}}, \bibinfo {author} {\bibfnamefont {M.}~\bibnamefont {Bibes}}, \
  and\ \bibinfo {author} {\bibfnamefont {A.}~\bibnamefont {Zunger}},\
  }\href@noop {} {\bibfield  {journal} {\bibinfo  {journal} {Phys. Rev. B}\
  }\textbf {\bibinfo {volume} {100}},\ \bibinfo {pages} {035119} (\bibinfo
  {year} {2019})}\BibitemShut {NoStop}%
\bibitem [{\citenamefont {Jana}, \citenamefont {Patra},\ and\ \citenamefont
  {Samal}(2018)}]{JanaJCP18a}%
  \BibitemOpen
  \bibfield  {author} {\bibinfo {author} {\bibfnamefont {S.}~\bibnamefont
  {Jana}}, \bibinfo {author} {\bibfnamefont {A.}~\bibnamefont {Patra}}, \ and\
  \bibinfo {author} {\bibfnamefont {P.}~\bibnamefont {Samal}},\ }\href@noop {}
  {\bibfield  {journal} {\bibinfo  {journal} {J. Chem. Phys.}\ }\textbf
  {\bibinfo {volume} {149}},\ \bibinfo {pages} {044120} (\bibinfo {year}
  {2018})}\BibitemShut {NoStop}%
\bibitem [{\citenamefont {Fu}\ and\ \citenamefont {Singh}(2018)}]{FuPRL18}%
  \BibitemOpen
  \bibfield  {author} {\bibinfo {author} {\bibfnamefont {Y.}~\bibnamefont
  {Fu}}\ and\ \bibinfo {author} {\bibfnamefont {D.~J.}\ \bibnamefont {Singh}},\
  }\href@noop {} {\bibfield  {journal} {\bibinfo  {journal} {Phys. Rev. Lett.}\
  }\textbf {\bibinfo {volume} {121}},\ \bibinfo {pages} {207201} (\bibinfo
  {year} {2018})}\BibitemShut {NoStop}%
\bibitem [{\citenamefont {Mej\'{\i}a-Rodr\'{\i}guez}\ and\ \citenamefont
  {Trickey}(2019)}]{MejiaRodriguezPRB19}%
  \BibitemOpen
  \bibfield  {author} {\bibinfo {author} {\bibfnamefont {D.}~\bibnamefont
  {Mej\'{\i}a-Rodr\'{\i}guez}}\ and\ \bibinfo {author} {\bibfnamefont {S.~B.}\
  \bibnamefont {Trickey}},\ }\href@noop {} {\bibfield  {journal} {\bibinfo
  {journal} {Phys. Rev. B}\ }\textbf {\bibinfo {volume} {100}},\ \bibinfo
  {pages} {041113(R)} (\bibinfo {year} {2019})}\BibitemShut {NoStop}%
\bibitem [{\citenamefont {Neumann}, \citenamefont {Nobes},\ and\ \citenamefont
  {Handy}(1996)}]{NeumannMP96}%
  \BibitemOpen
  \bibfield  {author} {\bibinfo {author} {\bibfnamefont {R.}~\bibnamefont
  {Neumann}}, \bibinfo {author} {\bibfnamefont {R.~H.}\ \bibnamefont {Nobes}},
  \ and\ \bibinfo {author} {\bibfnamefont {N.~C.}\ \bibnamefont {Handy}},\
  }\href@noop {} {\bibfield  {journal} {\bibinfo  {journal} {Mol. Phys.}\
  }\textbf {\bibinfo {volume} {87}},\ \bibinfo {pages} {1} (\bibinfo {year}
  {1996})}\BibitemShut {NoStop}%
\bibitem [{\citenamefont {Bienvenu}\ and\ \citenamefont
  {Knizia}(2018)}]{BienvenuJCTC18}%
  \BibitemOpen
  \bibfield  {author} {\bibinfo {author} {\bibfnamefont {A.~V.}\ \bibnamefont
  {Bienvenu}}\ and\ \bibinfo {author} {\bibfnamefont {G.}~\bibnamefont
  {Knizia}},\ }\href@noop {} {\bibfield  {journal} {\bibinfo  {journal} {J.
  Chem. Theory Comput.}\ }\textbf {\bibinfo {volume} {14}},\ \bibinfo {pages}
  {1297} (\bibinfo {year} {2018})}\BibitemShut {NoStop}%
\bibitem [{\citenamefont {Mejia-Rodriguez}\ and\ \citenamefont
  {Trickey}(2018)}]{MejiaRodriguezPRB18}%
  \BibitemOpen
  \bibfield  {author} {\bibinfo {author} {\bibfnamefont {D.}~\bibnamefont
  {Mejia-Rodriguez}}\ and\ \bibinfo {author} {\bibfnamefont {S.~B.}\
  \bibnamefont {Trickey}},\ }\href@noop {} {\bibfield  {journal} {\bibinfo
  {journal} {Phys. Rev. B}\ }\textbf {\bibinfo {volume} {98}},\ \bibinfo
  {pages} {115161} (\bibinfo {year} {2018})}\BibitemShut {NoStop}%
\bibitem [{\citenamefont {Morales-Garc\'{i}a}, \citenamefont {Valero},\ and\
  \citenamefont {Illas}(2017)}]{MoralesGarciaJPCC17}%
  \BibitemOpen
  \bibfield  {author} {\bibinfo {author} {\bibfnamefont {{\'A}.}~\bibnamefont
  {Morales-Garc\'{i}a}}, \bibinfo {author} {\bibfnamefont {R.}~\bibnamefont
  {Valero}}, \ and\ \bibinfo {author} {\bibfnamefont {F.}~\bibnamefont
  {Illas}},\ }\href@noop {} {\bibfield  {journal} {\bibinfo  {journal} {J.
  Phys. Chem. C}\ }\textbf {\bibinfo {volume} {121}},\ \bibinfo {pages} {18862}
  (\bibinfo {year} {2017})}\BibitemShut {NoStop}%
\bibitem [{\citenamefont {Chen}\ \emph {et~al.}(2018)\citenamefont {Chen},
  \citenamefont {Miceli}, \citenamefont {Rignanese},\ and\ \citenamefont
  {Pasquarello}}]{ChenPRM18}%
  \BibitemOpen
  \bibfield  {author} {\bibinfo {author} {\bibfnamefont {W.}~\bibnamefont
  {Chen}}, \bibinfo {author} {\bibfnamefont {G.}~\bibnamefont {Miceli}},
  \bibinfo {author} {\bibfnamefont {G.-M.}\ \bibnamefont {Rignanese}}, \ and\
  \bibinfo {author} {\bibfnamefont {A.}~\bibnamefont {Pasquarello}},\
  }\href@noop {} {\bibfield  {journal} {\bibinfo  {journal} {Phys. Rev.
  Materials}\ }\textbf {\bibinfo {volume} {2}},\ \bibinfo {pages} {073803}
  (\bibinfo {year} {2018})}\BibitemShut {NoStop}%
\bibitem [{\citenamefont {Trushin}\ \emph {et~al.}(2016)\citenamefont
  {Trushin}, \citenamefont {Betzinger}, \citenamefont {Bl\"ugel},\ and\
  \citenamefont {G\"orling}}]{TrushinPRB16}%
  \BibitemOpen
  \bibfield  {author} {\bibinfo {author} {\bibfnamefont {E.}~\bibnamefont
  {Trushin}}, \bibinfo {author} {\bibfnamefont {M.}~\bibnamefont {Betzinger}},
  \bibinfo {author} {\bibfnamefont {S.}~\bibnamefont {Bl\"ugel}}, \ and\
  \bibinfo {author} {\bibfnamefont {A.}~\bibnamefont {G\"orling}},\ }\href@noop
  {} {\bibfield  {journal} {\bibinfo  {journal} {Phys. Rev. B}\ }\textbf
  {\bibinfo {volume} {94}},\ \bibinfo {pages} {075123} (\bibinfo {year}
  {2016})}\BibitemShut {NoStop}%
\bibitem [{\citenamefont {Vl\v{c}ek}\ \emph {et~al.}(2015)\citenamefont
  {Vl\v{c}ek}, \citenamefont {Eisenberg}, \citenamefont {Steinle-Neumann},
  \citenamefont {Kronik},\ and\ \citenamefont {Baer}}]{VlcekJCP15}%
  \BibitemOpen
  \bibfield  {author} {\bibinfo {author} {\bibfnamefont {V.}~\bibnamefont
  {Vl\v{c}ek}}, \bibinfo {author} {\bibfnamefont {H.~R.}\ \bibnamefont
  {Eisenberg}}, \bibinfo {author} {\bibfnamefont {G.}~\bibnamefont
  {Steinle-Neumann}}, \bibinfo {author} {\bibfnamefont {L.}~\bibnamefont
  {Kronik}}, \ and\ \bibinfo {author} {\bibfnamefont {R.}~\bibnamefont
  {Baer}},\ }\href@noop {} {\bibfield  {journal} {\bibinfo  {journal} {J. Chem.
  Phys.}\ }\textbf {\bibinfo {volume} {142}},\ \bibinfo {pages} {034107}
  (\bibinfo {year} {2015})}\BibitemShut {NoStop}%
\bibitem [{\citenamefont {Pederson}, \citenamefont {Ruzsinszky},\ and\
  \citenamefont {Perdew}(2014)}]{PedersonJCP14}%
  \BibitemOpen
  \bibfield  {author} {\bibinfo {author} {\bibfnamefont {M.~R.}\ \bibnamefont
  {Pederson}}, \bibinfo {author} {\bibfnamefont {A.}~\bibnamefont
  {Ruzsinszky}}, \ and\ \bibinfo {author} {\bibfnamefont {J.~P.}\ \bibnamefont
  {Perdew}},\ }\href@noop {} {\bibfield  {journal} {\bibinfo  {journal} {J.
  Chem. Phys.}\ }\textbf {\bibinfo {volume} {140}},\ \bibinfo {pages} {121103}
  (\bibinfo {year} {2014})}\BibitemShut {NoStop}%
\bibitem [{\citenamefont {Tao}\ \emph {et~al.}(2003)\citenamefont {Tao},
  \citenamefont {Perdew}, \citenamefont {Staroverov},\ and\ \citenamefont
  {Scuseria}}]{TaoPRL03}%
  \BibitemOpen
  \bibfield  {author} {\bibinfo {author} {\bibfnamefont {J.}~\bibnamefont
  {Tao}}, \bibinfo {author} {\bibfnamefont {J.~P.}\ \bibnamefont {Perdew}},
  \bibinfo {author} {\bibfnamefont {V.~N.}\ \bibnamefont {Staroverov}}, \ and\
  \bibinfo {author} {\bibfnamefont {G.~E.}\ \bibnamefont {Scuseria}},\
  }\href@noop {} {\bibfield  {journal} {\bibinfo  {journal} {Phys. Rev. Lett.}\
  }\textbf {\bibinfo {volume} {91}},\ \bibinfo {pages} {146401} (\bibinfo
  {year} {2003})}\BibitemShut {NoStop}%
\bibitem [{\citenamefont {Perdew}\ \emph {et~al.}(2009)\citenamefont {Perdew},
  \citenamefont {Ruzsinszky}, \citenamefont {Csonka}, \citenamefont
  {Constantin},\ and\ \citenamefont {Sun}}]{PerdewPRL09}%
  \BibitemOpen
  \bibfield  {author} {\bibinfo {author} {\bibfnamefont {J.~P.}\ \bibnamefont
  {Perdew}}, \bibinfo {author} {\bibfnamefont {A.}~\bibnamefont {Ruzsinszky}},
  \bibinfo {author} {\bibfnamefont {G.~I.}\ \bibnamefont {Csonka}}, \bibinfo
  {author} {\bibfnamefont {L.~A.}\ \bibnamefont {Constantin}}, \ and\ \bibinfo
  {author} {\bibfnamefont {J.}~\bibnamefont {Sun}},\ }\href@noop {} {\bibfield
  {journal} {\bibinfo  {journal} {Phys. Rev. Lett.}\ }\textbf {\bibinfo
  {volume} {103}},\ \bibinfo {pages} {026403} (\bibinfo {year} {2009})},\
  \bibinfo {note} {\textbf{106}, 179902 (2011)}\BibitemShut {NoStop}%
\bibitem [{\citenamefont {Sun}, \citenamefont {Perdew},\ and\ \citenamefont
  {Ruzsinszky}(2015)}]{SunPNAS15}%
  \BibitemOpen
  \bibfield  {author} {\bibinfo {author} {\bibfnamefont {J.}~\bibnamefont
  {Sun}}, \bibinfo {author} {\bibfnamefont {J.~P.}\ \bibnamefont {Perdew}}, \
  and\ \bibinfo {author} {\bibfnamefont {A.}~\bibnamefont {Ruzsinszky}},\
  }\href@noop {} {\bibfield  {journal} {\bibinfo  {journal} {Proc. Natl. Acad.
  Sci. U.S.A.}\ }\textbf {\bibinfo {volume} {112}},\ \bibinfo {pages} {685}
  (\bibinfo {year} {2015})}\BibitemShut {NoStop}%
\bibitem [{\citenamefont {Bart\'{o}k}\ and\ \citenamefont
  {Yates}(2019)}]{BartokJCP19}%
  \BibitemOpen
  \bibfield  {author} {\bibinfo {author} {\bibfnamefont {A.~P.}\ \bibnamefont
  {Bart\'{o}k}}\ and\ \bibinfo {author} {\bibfnamefont {J.~R.}\ \bibnamefont
  {Yates}},\ }\href@noop {} {\bibfield  {journal} {\bibinfo  {journal} {J.
  Chem. Phys.}\ }\textbf {\bibinfo {volume} {150}},\ \bibinfo {pages} {161101}
  (\bibinfo {year} {2019})}\BibitemShut {NoStop}%
\bibitem [{\citenamefont {Tao}\ and\ \citenamefont {Mo}(2016)}]{TaoPRL16}%
  \BibitemOpen
  \bibfield  {author} {\bibinfo {author} {\bibfnamefont {J.}~\bibnamefont
  {Tao}}\ and\ \bibinfo {author} {\bibfnamefont {Y.}~\bibnamefont {Mo}},\
  }\href@noop {} {\bibfield  {journal} {\bibinfo  {journal} {Phys. Rev. Lett.}\
  }\textbf {\bibinfo {volume} {117}},\ \bibinfo {pages} {073001} (\bibinfo
  {year} {2016})}\BibitemShut {NoStop}%
\bibitem [{\citenamefont {Verma}\ and\ \citenamefont
  {Truhlar}(2017{\natexlab{b}})}]{VermaJPCC17}%
  \BibitemOpen
  \bibfield  {author} {\bibinfo {author} {\bibfnamefont {P.}~\bibnamefont
  {Verma}}\ and\ \bibinfo {author} {\bibfnamefont {D.~G.}\ \bibnamefont
  {Truhlar}},\ }\href@noop {} {\bibfield  {journal} {\bibinfo  {journal} {J.
  Phys. Chem. C}\ }\textbf {\bibinfo {volume} {121}},\ \bibinfo {pages} {7144}
  (\bibinfo {year} {2017}{\natexlab{b}})}\BibitemShut {NoStop}%
\bibitem [{\citenamefont {Hammer}, \citenamefont {Hansen},\ and\ \citenamefont
  {N{\o}rskov}(1999)}]{HammerPRB99}%
  \BibitemOpen
  \bibfield  {author} {\bibinfo {author} {\bibfnamefont {B.}~\bibnamefont
  {Hammer}}, \bibinfo {author} {\bibfnamefont {L.~B.}\ \bibnamefont {Hansen}},
  \ and\ \bibinfo {author} {\bibfnamefont {J.~K.}\ \bibnamefont {N{\o}rskov}},\
  }\href@noop {} {\bibfield  {journal} {\bibinfo  {journal} {Phys. Rev. B}\
  }\textbf {\bibinfo {volume} {59}},\ \bibinfo {pages} {7413} (\bibinfo {year}
  {1999})}\BibitemShut {NoStop}%
\bibitem [{\citenamefont {Perdew}\ \emph {et~al.}(2008)\citenamefont {Perdew},
  \citenamefont {Ruzsinszky}, \citenamefont {Csonka}, \citenamefont {Vydrov},
  \citenamefont {Scuseria}, \citenamefont {Constantin}, \citenamefont {Zhou},\
  and\ \citenamefont {Burke}}]{PerdewPRL08}%
  \BibitemOpen
  \bibfield  {author} {\bibinfo {author} {\bibfnamefont {J.~P.}\ \bibnamefont
  {Perdew}}, \bibinfo {author} {\bibfnamefont {A.}~\bibnamefont {Ruzsinszky}},
  \bibinfo {author} {\bibfnamefont {G.~I.}\ \bibnamefont {Csonka}}, \bibinfo
  {author} {\bibfnamefont {O.~A.}\ \bibnamefont {Vydrov}}, \bibinfo {author}
  {\bibfnamefont {G.~E.}\ \bibnamefont {Scuseria}}, \bibinfo {author}
  {\bibfnamefont {L.~A.}\ \bibnamefont {Constantin}}, \bibinfo {author}
  {\bibfnamefont {X.}~\bibnamefont {Zhou}}, \ and\ \bibinfo {author}
  {\bibfnamefont {K.}~\bibnamefont {Burke}},\ }\href@noop {} {\bibfield
  {journal} {\bibinfo  {journal} {Phys. Rev. Lett.}\ }\textbf {\bibinfo
  {volume} {100}},\ \bibinfo {pages} {136406} (\bibinfo {year} {2008})},\
  \bibinfo {note} {\textbf{102}, 039902(E) (2009)}\BibitemShut {NoStop}%
\bibitem [{\citenamefont {Engel}\ and\ \citenamefont
  {Vosko}(1993)}]{EngelPRB93}%
  \BibitemOpen
  \bibfield  {author} {\bibinfo {author} {\bibfnamefont {E.}~\bibnamefont
  {Engel}}\ and\ \bibinfo {author} {\bibfnamefont {S.~H.}\ \bibnamefont
  {Vosko}},\ }\href@noop {} {\bibfield  {journal} {\bibinfo  {journal} {Phys.
  Rev. B}\ }\textbf {\bibinfo {volume} {47}},\ \bibinfo {pages} {13164}
  (\bibinfo {year} {1993})}\BibitemShut {NoStop}%
\bibitem [{\citenamefont {Perdew}\ \emph {et~al.}(1992)\citenamefont {Perdew},
  \citenamefont {Chevary}, \citenamefont {Vosko}, \citenamefont {Jackson},
  \citenamefont {Pederson}, \citenamefont {Singh},\ and\ \citenamefont
  {Fiolhais}}]{PerdewPRB92b}%
  \BibitemOpen
  \bibfield  {author} {\bibinfo {author} {\bibfnamefont {J.~P.}\ \bibnamefont
  {Perdew}}, \bibinfo {author} {\bibfnamefont {J.~A.}\ \bibnamefont {Chevary}},
  \bibinfo {author} {\bibfnamefont {S.~H.}\ \bibnamefont {Vosko}}, \bibinfo
  {author} {\bibfnamefont {K.~A.}\ \bibnamefont {Jackson}}, \bibinfo {author}
  {\bibfnamefont {M.~R.}\ \bibnamefont {Pederson}}, \bibinfo {author}
  {\bibfnamefont {D.~J.}\ \bibnamefont {Singh}}, \ and\ \bibinfo {author}
  {\bibfnamefont {C.}~\bibnamefont {Fiolhais}},\ }\href@noop {} {\bibfield
  {journal} {\bibinfo  {journal} {Phys. Rev. B}\ }\textbf {\bibinfo {volume}
  {46}},\ \bibinfo {pages} {6671} (\bibinfo {year} {1992})},\ \bibinfo {note}
  {\textbf{48}, 4978(E) (1993)}\BibitemShut {NoStop}%
\bibitem [{\citenamefont {Boese}\ and\ \citenamefont
  {Handy}(2001)}]{BoeseJCP01}%
  \BibitemOpen
  \bibfield  {author} {\bibinfo {author} {\bibfnamefont {A.~D.}\ \bibnamefont
  {Boese}}\ and\ \bibinfo {author} {\bibfnamefont {N.~C.}\ \bibnamefont
  {Handy}},\ }\href@noop {} {\bibfield  {journal} {\bibinfo  {journal} {J.
  Chem. Phys.}\ }\textbf {\bibinfo {volume} {114}},\ \bibinfo {pages} {5497}
  (\bibinfo {year} {2001})}\BibitemShut {NoStop}%
\bibitem [{\citenamefont {Perdew}\ and\ \citenamefont
  {Wang}(1992)}]{PerdewPRB92a}%
  \BibitemOpen
  \bibfield  {author} {\bibinfo {author} {\bibfnamefont {J.~P.}\ \bibnamefont
  {Perdew}}\ and\ \bibinfo {author} {\bibfnamefont {Y.}~\bibnamefont {Wang}},\
  }\href@noop {} {\bibfield  {journal} {\bibinfo  {journal} {Phys. Rev. B}\
  }\textbf {\bibinfo {volume} {45}},\ \bibinfo {pages} {13244} (\bibinfo {year}
  {1992})},\ \bibinfo {note} {\textbf{98}, 079904(E) (2018)}\BibitemShut
  {NoStop}%
\bibitem [{\citenamefont {van Leeuwen}\ and\ \citenamefont
  {Baerends}(1994)}]{vanLeeuwenPRA94}%
  \BibitemOpen
  \bibfield  {author} {\bibinfo {author} {\bibfnamefont {R.}~\bibnamefont {van
  Leeuwen}}\ and\ \bibinfo {author} {\bibfnamefont {E.~J.}\ \bibnamefont
  {Baerends}},\ }\href@noop {} {\bibfield  {journal} {\bibinfo  {journal}
  {Phys. Rev. A}\ }\textbf {\bibinfo {volume} {49}},\ \bibinfo {pages} {2421}
  (\bibinfo {year} {1994})}\BibitemShut {NoStop}%
\bibitem [{\citenamefont {Finzel}\ and\ \citenamefont
  {Baranov}(2017)}]{FinzelIJQC17}%
  \BibitemOpen
  \bibfield  {author} {\bibinfo {author} {\bibfnamefont {K.}~\bibnamefont
  {Finzel}}\ and\ \bibinfo {author} {\bibfnamefont {A.~I.}\ \bibnamefont
  {Baranov}},\ }\href@noop {} {\bibfield  {journal} {\bibinfo  {journal} {Int.
  J. Quantum Chem.}\ }\textbf {\bibinfo {volume} {117}},\ \bibinfo {pages} {40}
  (\bibinfo {year} {2017})}\BibitemShut {NoStop}%
\bibitem [{\citenamefont {Blaha}\ \emph {et~al.}(2018)\citenamefont {Blaha},
  \citenamefont {Schwarz}, \citenamefont {Madsen}, \citenamefont {Kvasnicka},
  \citenamefont {Luitz}, \citenamefont {Laskowski}, \citenamefont {Tran},\ and\
  \citenamefont {Marks}}]{WIEN2k}%
  \BibitemOpen
  \bibfield  {author} {\bibinfo {author} {\bibfnamefont {P.}~\bibnamefont
  {Blaha}}, \bibinfo {author} {\bibfnamefont {K.}~\bibnamefont {Schwarz}},
  \bibinfo {author} {\bibfnamefont {G.~K.~H.}\ \bibnamefont {Madsen}}, \bibinfo
  {author} {\bibfnamefont {D.}~\bibnamefont {Kvasnicka}}, \bibinfo {author}
  {\bibfnamefont {J.}~\bibnamefont {Luitz}}, \bibinfo {author} {\bibfnamefont
  {R.}~\bibnamefont {Laskowski}}, \bibinfo {author} {\bibfnamefont
  {F.}~\bibnamefont {Tran}}, \ and\ \bibinfo {author} {\bibfnamefont {L.~D.}\
  \bibnamefont {Marks}},\ }\href@noop {} {\emph {\bibinfo {title} {WIEN2k: An
  Augmented Plane Wave plus Local Orbitals Program for Calculating Crystal
  Properties}}}\ (\bibinfo  {publisher} {Vienna University of Technology},\
  \bibinfo {address} {Austria},\ \bibinfo {year} {2018})\BibitemShut {NoStop}%
\bibitem [{\citenamefont {Singh}\ and\ \citenamefont
  {Nordstr{\"{o}}m}(2006)}]{Singh}%
  \BibitemOpen
  \bibfield  {author} {\bibinfo {author} {\bibfnamefont {D.~J.}\ \bibnamefont
  {Singh}}\ and\ \bibinfo {author} {\bibfnamefont {L.}~\bibnamefont
  {Nordstr{\"{o}}m}},\ }\href@noop {} {\emph {\bibinfo {title} {Planewaves,
  Pseudopotentials, and the LAPW Method, 2nd ed.}}}\ (\bibinfo  {publisher}
  {Springer},\ \bibinfo {address} {New York},\ \bibinfo {year}
  {2006})\BibitemShut {NoStop}%
\bibitem [{\citenamefont {Karsai}, \citenamefont {Tran},\ and\ \citenamefont
  {Blaha}(2017)}]{KarsaiCPC17}%
  \BibitemOpen
  \bibfield  {author} {\bibinfo {author} {\bibfnamefont {F.}~\bibnamefont
  {Karsai}}, \bibinfo {author} {\bibfnamefont {F.}~\bibnamefont {Tran}}, \ and\
  \bibinfo {author} {\bibfnamefont {P.}~\bibnamefont {Blaha}},\ }\href@noop {}
  {\bibfield  {journal} {\bibinfo  {journal} {Comput. Phys. Commun.}\ }\textbf
  {\bibinfo {volume} {220}},\ \bibinfo {pages} {230} (\bibinfo {year}
  {2017})}\BibitemShut {NoStop}%
\bibitem [{\citenamefont {Borlido}\ \emph {et~al.}(2019)\citenamefont
  {Borlido}, \citenamefont {Aull}, \citenamefont {Huran}, \citenamefont {Tran},
  \citenamefont {Marques},\ and\ \citenamefont {Botti}}]{BorlidoJCTC19}%
  \BibitemOpen
  \bibfield  {author} {\bibinfo {author} {\bibfnamefont {P.}~\bibnamefont
  {Borlido}}, \bibinfo {author} {\bibfnamefont {T.}~\bibnamefont {Aull}},
  \bibinfo {author} {\bibfnamefont {A.~W.}\ \bibnamefont {Huran}}, \bibinfo
  {author} {\bibfnamefont {F.}~\bibnamefont {Tran}}, \bibinfo {author}
  {\bibfnamefont {M.~A.~L.}\ \bibnamefont {Marques}}, \ and\ \bibinfo {author}
  {\bibfnamefont {S.}~\bibnamefont {Botti}},\ }\href@noop {} {\bibfield
  {journal} {\bibinfo  {journal} {J. Chem. Theory Comput.}\ }\textbf {\bibinfo
  {volume} {15}},\ \bibinfo {pages} {5069} (\bibinfo {year}
  {2019})}\BibitemShut {NoStop}%
\bibitem [{\citenamefont {Kresse}\ and\ \citenamefont
  {Furthm\"uller}(1996)}]{KressePRB96}%
  \BibitemOpen
  \bibfield  {author} {\bibinfo {author} {\bibfnamefont {G.}~\bibnamefont
  {Kresse}}\ and\ \bibinfo {author} {\bibfnamefont {J.}~\bibnamefont
  {Furthm\"uller}},\ }\href@noop {} {\bibfield  {journal} {\bibinfo  {journal}
  {Phys. Rev. B}\ }\textbf {\bibinfo {volume} {54}},\ \bibinfo {pages} {11169}
  (\bibinfo {year} {1996})}\BibitemShut {NoStop}%
\bibitem [{\citenamefont {Bl\"{o}chl}(1994)}]{BlochlPRB94b}%
  \BibitemOpen
  \bibfield  {author} {\bibinfo {author} {\bibfnamefont {P.~E.}\ \bibnamefont
  {Bl\"{o}chl}},\ }\href@noop {} {\bibfield  {journal} {\bibinfo  {journal}
  {Phys. Rev. B}\ }\textbf {\bibinfo {volume} {50}},\ \bibinfo {pages} {17953}
  (\bibinfo {year} {1994})}\BibitemShut {NoStop}%
\bibitem [{\citenamefont {Payne}\ \emph {et~al.}(1992)\citenamefont {Payne},
  \citenamefont {Teter}, \citenamefont {Allan}, \citenamefont {Arias},\ and\
  \citenamefont {Joannopoulos}}]{PayneRMP92}%
  \BibitemOpen
  \bibfield  {author} {\bibinfo {author} {\bibfnamefont {M.~C.}\ \bibnamefont
  {Payne}}, \bibinfo {author} {\bibfnamefont {M.~P.}\ \bibnamefont {Teter}},
  \bibinfo {author} {\bibfnamefont {D.~C.}\ \bibnamefont {Allan}}, \bibinfo
  {author} {\bibfnamefont {T.~A.}\ \bibnamefont {Arias}}, \ and\ \bibinfo
  {author} {\bibfnamefont {J.~D.}\ \bibnamefont {Joannopoulos}},\ }\href@noop
  {} {\bibfield  {journal} {\bibinfo  {journal} {Rev. Mod. Phys.}\ }\textbf
  {\bibinfo {volume} {64}},\ \bibinfo {pages} {1045} (\bibinfo {year}
  {1992})}\BibitemShut {NoStop}%
\bibitem [{\citenamefont {Clark}\ \emph {et~al.}(2005)\citenamefont {Clark},
  \citenamefont {Segall}, \citenamefont {Pickard}, \citenamefont {Hasnip},
  \citenamefont {Probert}, \citenamefont {Refson},\ and\ \citenamefont
  {Payne}}]{ClarkZKCM05}%
  \BibitemOpen
  \bibfield  {author} {\bibinfo {author} {\bibfnamefont {S.~J.}\ \bibnamefont
  {Clark}}, \bibinfo {author} {\bibfnamefont {M.~D.}\ \bibnamefont {Segall}},
  \bibinfo {author} {\bibfnamefont {C.~J.}\ \bibnamefont {Pickard}}, \bibinfo
  {author} {\bibfnamefont {P.~J.}\ \bibnamefont {Hasnip}}, \bibinfo {author}
  {\bibfnamefont {M.~I.~J.}\ \bibnamefont {Probert}}, \bibinfo {author}
  {\bibfnamefont {K.}~\bibnamefont {Refson}}, \ and\ \bibinfo {author}
  {\bibfnamefont {M.~C.}\ \bibnamefont {Payne}},\ }\href@noop {} {\bibfield
  {journal} {\bibinfo  {journal} {Z. Kristallogr.-Cryst. Mater.}\ }\textbf
  {\bibinfo {volume} {220}},\ \bibinfo {pages} {567} (\bibinfo {year}
  {2005})}\BibitemShut {NoStop}%
\bibitem [{\citenamefont {Lima}\ \emph {et~al.}(2007)\citenamefont {Lima},
  \citenamefont {Pedroza}, \citenamefont {da~Silva}, \citenamefont {Fazzio},
  \citenamefont {Vieira}, \citenamefont {Freire},\ and\ \citenamefont
  {Capelle}}]{LimaJCP07}%
  \BibitemOpen
  \bibfield  {author} {\bibinfo {author} {\bibfnamefont {M.~P.}\ \bibnamefont
  {Lima}}, \bibinfo {author} {\bibfnamefont {L.~S.}\ \bibnamefont {Pedroza}},
  \bibinfo {author} {\bibfnamefont {A.~J.~R.}\ \bibnamefont {da~Silva}},
  \bibinfo {author} {\bibfnamefont {A.}~\bibnamefont {Fazzio}}, \bibinfo
  {author} {\bibfnamefont {D.}~\bibnamefont {Vieira}}, \bibinfo {author}
  {\bibfnamefont {H.~J.~P.}\ \bibnamefont {Freire}}, \ and\ \bibinfo {author}
  {\bibfnamefont {K.}~\bibnamefont {Capelle}},\ }\href@noop {} {\bibfield
  {journal} {\bibinfo  {journal} {J. Chem. Phys.}\ }\textbf {\bibinfo {volume}
  {126}},\ \bibinfo {pages} {144107} (\bibinfo {year} {2007})}\BibitemShut
  {NoStop}%
\end{thebibliography}%

\end{document}